\documentclass[letters,fleqn,usenatbib]{mnras}

\usepackage[T1]{fontenc}
\usepackage{ae,aecompl}
\usepackage{graphicx}	      
\usepackage{amsmath}	      
\usepackage{amssymb}	      
\usepackage[normalem]{ulem}   

\newcommand {\eV}          {\,\rm eV}

\newcommand {\Mpc}         {\,\rm Mpc}
\newcommand {\Mpch}        {\,h^{-1}\,\rm Mpc}
\newcommand {\Msun}        {\, \rm{M}_{\sun}}
\newcommand {\psiDM}       {\psi \rm{DM}}
\newcommand {\FP}          {{\rm FP} \psi {\rm DM}}
\newcommand {\EA}          {{\rm EA} \psi {\rm DM}}
\newcommand {\initangle}   {\delta \theta_0}
\newcommand {\kJeq}        {k_{J,{\rm eq}}}
\newcommand {\PM}          {m_{\rm 22}}
\newcommand {\PMeff}       {m_{\rm 22, eff}}
\newcommand {\Mvir}        {M_{\rm h}}
\newcommand {\Nmajor}      {N_{\rm major, z57}}

\newcommand {\sref}[1]     {Section~\ref{#1}}
\newcommand {\fref}[1]     {Fig.~\ref{#1}}

\newcommand {\be}          {\begin{equation}}
\newcommand {\ee}          {\end{equation}}

\title[Halo Formation with $\EA$]{Halo Abundance and Assembly History with Extreme-Axion Wave Dark Matter at $z\ge 4$}

\author[Schive \& Chiueh]{
Hsi-Yu Schive,$^{1}$\thanks{E-mail: hyschive@ncsa.illinois.edu}
Tzihong Chiueh$^{2,3,4}$
\\
$^{1}$National Center for Supercomputing Applications, Urbana, IL, 61801, USA\\
$^{2}$Department of Physics, National Taiwan University, 10617 Taipei, Taiwan\\
$^{3}$Institute of Astrophysics, National Taiwan University, 10617 Taipei, Taiwan\\
$^{4}$Center for Theoretical Sciences, National Taiwan University, 10617 Taipei, Taiwan
}

\date{Accepted XXX. Received YYY; in original form ZZZ}

\pubyear{2017}

\begin{document}
\label{firstpage}
\pagerange{\pageref{firstpage}--\pageref{lastpage}}
\maketitle

\begin{abstract}

Wave dark matter ($\psiDM$) composed of extremely light bosons
($m_{\psi} \sim 10^{-22} \eV$), with quantum pressure suppressing
structures below a kpc-scale de Broglie wavelength, has become a viable
dark matter candidate.
Compared to the conventional free-particle $\psiDM$ ($\FP$), the
extreme-axion $\psiDM$ model ($\EA$) proposed by Zhang \& Chiueh (2017)
features a larger cut-off wavenumber and a broad spectral bump in the matter
transfer function.
Here we conduct cosmological simulations to compare the halo abundances
and assembly histories at $z=4 \text{--} 11$ between three different scenarios:
$\FP$, $\EA$, and cold dark matter (CDM).
We show that $\EA$ produces significantly more abundant low-mass haloes than
$\FP$ with the same $m_{\psi}$, and therefore could alleviate the tension in
$m_{\psi}$ required by the Ly$\alpha$ forest data and by the kpc-scale dwarf
galaxy cores.
We also find that, compared to the CDM counterparts, massive $\EA$ haloes are
on average $3\text{--}4$ times more massive at $z=10\text{--}11$ due to their
earlier formation, undergo a slower mass accretion at $7 \lesssim z \lesssim 11$,
and then show a rapidly rising major merger rate exceeding CDM by
$\sim 50\%$ at $4 \lesssim z \lesssim 7$.
This fact suggests that $\EA$ haloes may exhibit more prominent starbursts at $z \lesssim 7$.
\end{abstract}


\section{Introduction}
\label{sec:intro}

Wave dark matter \citep[$\psiDM$,][]{Hu2000, Schive2014a}
has become a promising dark matter candidate.
In this model, the dark matter is assumed to be composed of extremely light bosons with a
mass of $m_{\psi} \sim 10^{-22} \eV$, where the uncertainty principle leads to
a quantum pressure suppressing cosmic structures below the kpc scale. It thus
provides a plausible solution to the small-scale issues found in the dissipationless
cold dark matter (CDM) simulations \citep{Weinberg2013}. See \citet{Marsh2016}
and \citet{Hui2017} for comprehensive reviews on $\psiDM$.

The axion model is a $\psiDM$ candidate, for which the
field potential is specified by a cosine potential,
$V(\phi) = m_{\psi}^2 f^2(1-\cos(\phi/f))$, where $f$ is the axion decay constant and $\theta \equiv \phi/f$ is the axion angle.
In the very early epoch of radiation era, the initial angle of the background
field, $\theta_0 \equiv \pi-\initangle$, is frozen to $0 \le \theta_0 \le \pi$.
After the Compton length of the particle enters the horizon, the background
field starts to oscillate in the cosine potential. This is a damped oscillation
due to Hubble friction. Therefore, the angle rapidly sinks to the bottom of
the potential ($\theta \ll 1$) and samples the harmonic oscillator potential,
from which point on axions practically become free particles.
The free-particle $\psiDM$ ($\FP$) model in this context assumes
$\theta_0 \ll 1$ and therefore the field executes simple harmonic
oscillation right from the beginning \citep{ZC2017a}.

It turns out that perturbations of damped oscillations with different degrees of
nonlinearity have little difference in their spectra evaluated at the
radiation-matter equality, except for the extreme case where $\initangle \ll 1$
(i.e., $\theta_0 \sim \pi$), the extreme-axion $\psiDM$ ($\EA$) model
\citep{ZC2017b}. Here the initial field is located near the unstable
equilibrium on the potential top, and hence the oscillation can be
substantially delayed compared with other initial angles. The delay weakens
the role of Hubble friction and excites a parametric instability arising from
small residual nonlinearity in the oscillation. This instability is weak but
gets stronger when $\initangle \to 0$. In this paper we take
$\initangle = 0.2^{\circ}$ as a representative example,
corresponding to $4\pi Gf^2=1.13\times10^{-5}$ where $G$ is the gravitational constant.

Generally speaking, there are two categories of constraints on the $\psiDM$
particle mass, $\PM \equiv m_{\psi}/10^{-22} \eV$. The first class of
constraints addresses the kpc-scale cores found in the dwarf spheroidal galaxies,
leading to $\PM \sim 1$ (e.g., \citet{Schive2014a, Chen2017}; see also \citet{Gonzalez-Morales2017}
who derived a smaller mass when considering the mass-anisotropy degeneracy) since the larger
the particle mass, the smaller the core. The second class of constraints focuses
on the abundance and size of cosmic small-scale structures, especially at
higher redshifts. For example, the Ly$\alpha$ forest data suggest $\PM \gtrsim 10$
\citep{Irsic2017, Armengaud2017}, and the high-$z$ luminosity functions
and reionization imply $\PM \gtrsim 1$ \cite[e.g.,][]{Schive2016, Corasaniti2017}.
There is thus a moderate but distinct tension between the two types of
constraints, which is similar to, but not as severe as, the ``Catch 22'' problem
of warm dark matter \citep{Maccio2012}. As found by \citet{ZC2017a,ZC2017b}, $\EA$
predicts significantly more abundant small-scale structures and therefore low-mass
haloes compared to $\FP$ with the same
particle mass, and thus could alleviate this tension. Here we report the
first quantitative study on this subject.

This paper is structured as follows. We describe the simulation setup in
\sref{sec:simu} and show the results of halo mass function and assembly history
in \sref{sec:results}. Finally, we discuss and summarize our findings in
\sref{sec:discussion}.

\section{Simulations}
\label{sec:simu}

\subsection{Initial Power Spectra}
\label{subsec:power_spec}
In the $\psiDM$ scenario, quantum pressure resulting from the uncertainty
principle suppresses the small-scale structures below a characteristic Jeans
scale. This suppression can be expressed by the CDM-to-$\psiDM$
transfer function, $T_{\psiDM}^2(k,z) = P_{\psiDM}(k,z) / P_{\rm CDM}(k,z)$,
where $P$ is the power spectrum. Although $T_{\psiDM}(k,z)$ is
in general redshift-dependent, \citet{Schive2016} showed that it can be well
approximated as redshift-independent for the particle masses ($\PM \sim 1$),
redshifts ($z\sim \text{4--11}$), and halo masses ($\Mvir \gtrsim 10^9 \Msun$)
relevant to this work, mainly because the Jeans mass is well below $10^9 \Msun$.

The $\FP$ transfer function is given by \citep{Hu2000}
\be
T_{\FP}(x) \approx \frac{\cos x^3}{1+x^8},\;\; x=1.61\,\PM^{1/18}\frac{k}{\kJeq},
\label{eq:TranFuncHu}
\ee
where $\kJeq=9\,\PM^{1/2}\Mpc^{-1}$ is the Jeans wavenumber at the
matter-radiation equality. It exhibits a sharp cut-off at $k \sim \kJeq$
and strong oscillations for $k > \kJeq$. In comparison,
the $\EA$ transfer function with the same particle mass
features a larger cut-off wavenumber and a spectral bump before the
cut-off \citep[see Fig. 3 in][]{ZC2017b}, which has been verified later by \citet{Cedeno2017}.

\begin{figure}
\centering
\includegraphics[width=7cm]{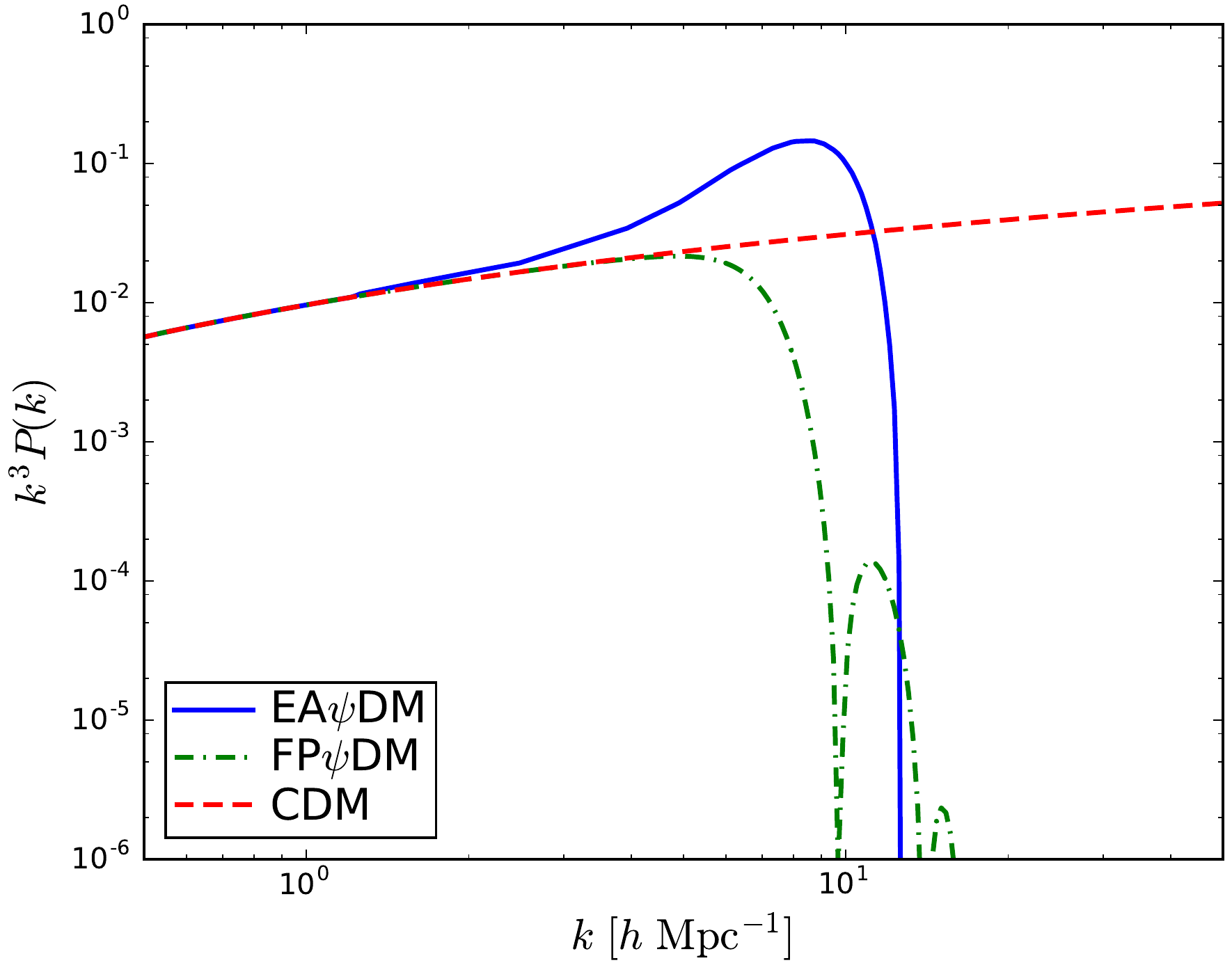}
\caption{
Linear power spectra of CDM, $\FP$, and $\EA$ at $z=100$.
Both $\psiDM$ power spectra feature a strong suppression at the high-$k$ end,
while $\EA$ shows a broad spectral bump peaking at $k \sim 8\,h\,\Mpc^{-1}$
and a cut-off wavenumber roughly twice larger than that of $\FP$.
}
\label{fig:power_spec}
\end{figure}

\fref{fig:power_spec} compares the linear matter power spectra of CDM,
$\FP$, and $\EA$ at $z=100$. We adopt the fiducial cosmological parameters of
$\Omega_{m}=0.30$, $\Omega_{\Lambda}=0.70$, $\Omega_{b}=0.06$, $h=0.70$, and
$\psiDM$ parameters of $\PM=1.1$ and $\initangle=0.2^{\circ}$.
The $\EA$ power spectrum shows a broad spectral bump peaking at
$k \sim 8\,h\,\Mpc^{-1}$ and exceeding the CDM power spectrum by a factor of
five, suggesting a significant excess of haloes with $\Mvir \sim 3\times10^{10} \Msun$.
In addition, the $\EA$ power spectrum exhibits a cut-off wavenumber about a
factor of two larger than that of $\FP$, indicative of significantly more haloes below
$\sim 10^{10} \Msun$. We quantify these differences from cosmological
simulations in \sref{sec:results}.

\subsection{Simulation Setup}
\label{subsec:simu_setup}

Genuine $\psiDM$ simulations solving the Schr\"{o}dinger-Poisson equation
are extremely time-consuming since the matter wave dispersion relation
demands exceptionally high spatial and temporal resolutions to
resolve the wave function accurately \citep[e.g.,][]{Schive2014a, Schive2014b}.
However, \citet{Schive2016} shows that collisionless $N$-body simulations
with $\psiDM$ initial power spectra can be adopted to study the $\psiDM$
evolution as long as the dynamical effect of quantum pressure is negligible
for the redshifts and halo masses of interest.
In this work, we focus on haloes more massive than
$\sim 2\times10^9 \Msun$, an order of magnitude higher than the $\psiDM$ Jeans
mass with $\PM \sim 1$ at $z \sim 10$, and we do not address the internal structure
of haloes (e.g., the cuspy or cored density profiles).
Therefore, it is sufficient for our purpose to conduct the collisionless $N$-body simulations.

We use the \texttt{CAMB} package \citep{Lewis2000} for generating the CDM
transfer function, the \texttt{MUSIC} code \citep{Hahn2011} for constructing
the initial conditions, and the \texttt{GADGET-2} code \citep{Springel2005}
for the $N$-body simulations. We adopt a fiducial simulation configuration of
$(L,\,N)=(80\Mpch,\,1024^3)$, where $L^3$ is the comoving box size and $N$ is the
total number of simulation particles. It corresponds to a particle mass resolution
of $\sim 5.7\times10^7 \Msun$. This configuration is chosen to both
accommodate a sufficient number of haloes above $\sim 10^{12} \Msun$ at $z \sim 4$
and to capture the decline of $\psiDM$ halo mass function
below $\sim 10^{10} \Msun$. We also conduct simulations with $(L,\,N)=(50\Mpch,\,1024^3)$
and $(160\Mpch,\,1024^3)$ to validate the numerical convergence. For each configuration,
we conduct CDM, $\FP$, and $\EA$ simulations from $z=100$ to $4$.

\section{Results}
\label{sec:results}

\begin{figure*}
\centering
\includegraphics[width=16cm]{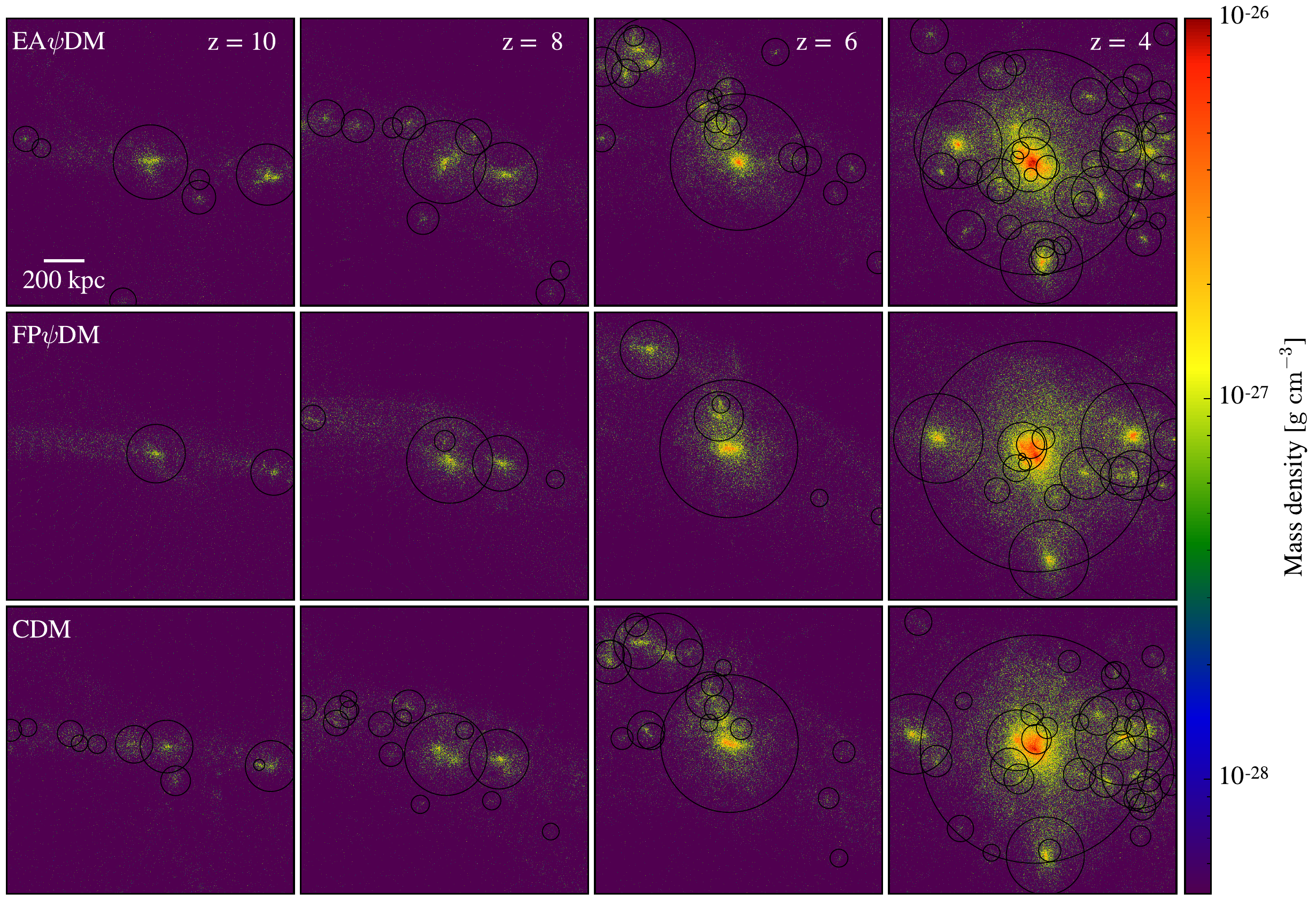}
\caption{
Projected dark matter density in a 1 $\Mpch$ thick slab centered on a
representative massive halo in the simulations. Different rows represent
different dark matter models, and different columns represent different
redshifts. Circles depict the halo virial radii. This halo has a similar
mass of $\Mvir \sim 6.5\times10^{12} \Msun$ at $z=4$ in all three models
but is apparently more massive in $\EA$ at $z=10$. In addition, $\FP$ shows
significantly fewer low-mass haloes at all redshifts. These facts suggest very
different halo formation histories in different models. See text for details.
The images are produced with the analysis toolkit \texttt{yt} \citep{Turk2011}.
}
\label{fig:proj_dens}
\end{figure*}

\fref{fig:proj_dens} shows the projected dark matter density centered on one of
the most massive haloes in the simulations at $z=\text{4--10}$, which is
unambiguously identified in all three models. At $z=10$, the $\EA$ halo has a
mass of $\Mvir \sim 2.3\times10^{11} \Msun$, about two and three times more
massive than the $\FP$ and CDM counterparts, respectively. However,
at $z=4$, the halo masses in different models converge to
$\Mvir \sim (\text{6.3--6.7})\times10^{12} \Msun$. Furthermore, $\FP$ shows
significantly fewer low-mass haloes at all redshifts. These facts suggest very
different halo mass functions and assembly histories in different
models, particularly at higher redshifts. We provide quantitative analyses in
this section.

\subsection{Halo Mass Function}
\label{subsec:mass_func}

\begin{figure}
\centering
\includegraphics[width=\columnwidth]{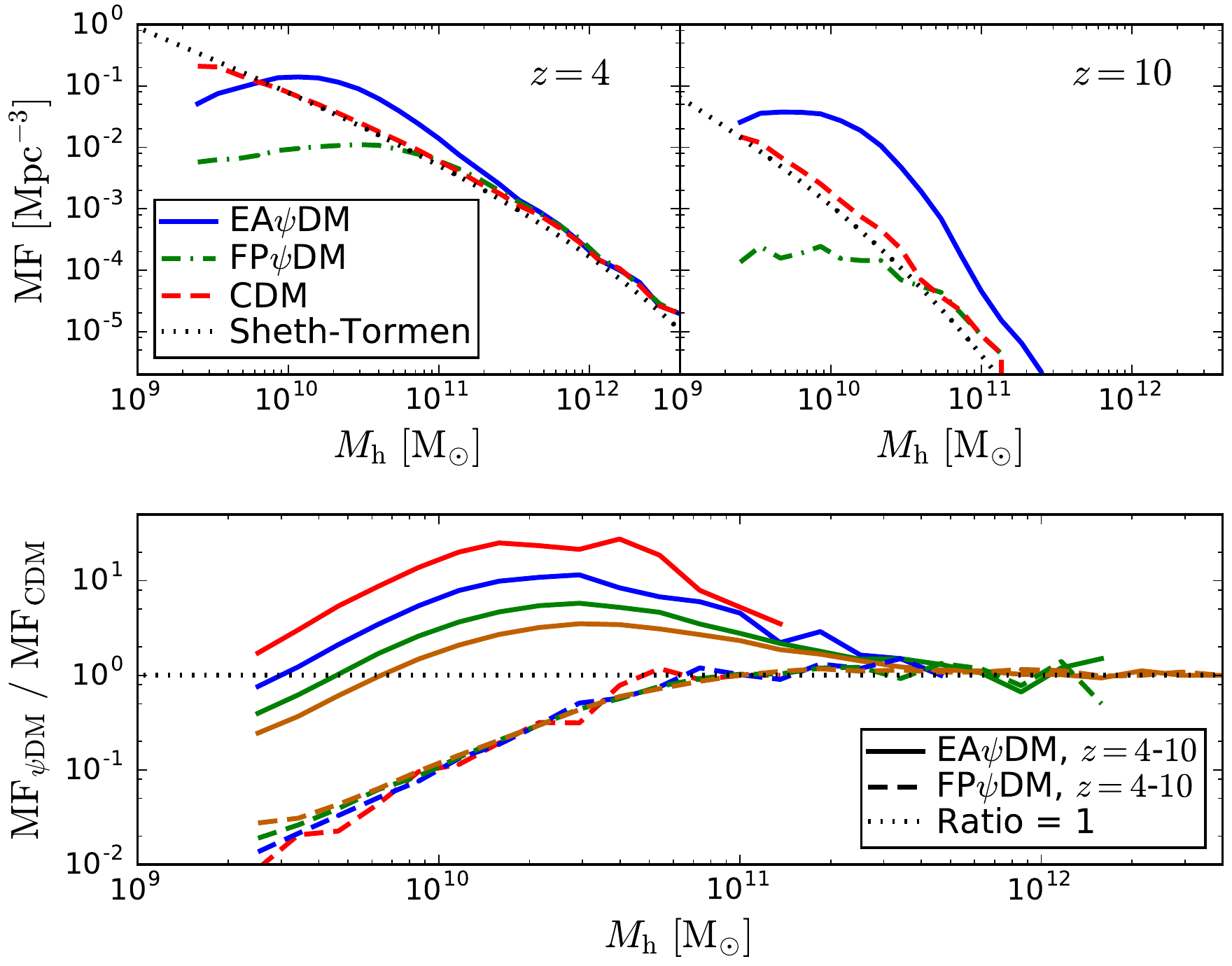}
\caption{
Halo mass functions (MF) in logarithmic mass bins. The upper panels show the
MFs at $z=4$ and $10$, where the dotted lines
represent the analytical prediction of \citet{Sheth1999}
for CDM. The lower panel shows the MF ratios between $\psiDM$ and CDM at
$z=$ 10 (red), 8 (blue), 6 (green), and 4 (brown). $\EA$ MF outnumbers CDM in a
wide mass range from several times $10^9 \Msun$ to several times $10^{11} \Msun$,
especially at higher redshifts. In comparison, $\FP$ has significantly fewer
low-mass haloes. See text for details.
}
\label{fig:mass_func}
\end{figure}

\fref{fig:mass_func} shows the halo mass function (MF) at $z=\text{4--10}$
and the MF ratios between different models.
We use the \texttt{AMIGA} Halo Finder \citep{Knollmann2009} to
identify all haloes with at least 40 particles, corresponding to a minimum halo
mass of $\sim 2.3\times10^{9} \Msun$ in a $L=80\Mpch$ box.
To validate the results, we compare the simulated CDM MF to the
analytical prediction of \citet{Sheth1999} and demonstrate a good agreement,
especially at lower redshifts. We also verify that the MFs shown
in \fref{fig:mass_func}, particularly at the low- and high-mass ends, are
consistent with simulations with $L=50\Mpch$ and $160\Mpch$.

The $\EA$ MF is found to
outnumber CDM in a wide mass range from several times $10^9 \Msun$
to several times $10^{11} \Msun$, with the maximum difference at
$\sim 3\times10^{10} \Msun$. It is consistent with the $\EA$ power spectrum
with a broad spectral bump peaking at $k \sim 8\,h\,\Mpc^{-1}$ (see \fref{fig:power_spec}).
The excess of the $\EA$ MF is more prominent at higher
redshifts, reaching a factor of $\sim \text{10--30}$ higher than CDM for
$\Mvir \sim 10^{10} \text{--} 10^{11} \Msun$ at $z=10$.
By contrast to the CDM haloes with similar masses, these abundant
high-$z$ $\EA$ haloes are the first collapsed objects which
accrete mass mainly by smooth mass accretion due to the strong suppression
of low-mass haloes and form earlier because of the higher local overdensity.

The difference between the $\EA$ and CDM MFs above $\sim 10^{10} \Msun$ diminishes
at lower redshifts but is still about two- to threefold for
$\Mvir \sim 10^{10} \text{--} 10^{11} \Msun$ at $z=4$. Moreover, $\EA$ has
substantially more haloes than $\FP$ even below $\sim 10^{10} \Msun$, although
both simulations adopt $\PM=1.1$. These unique features in $\EA$ may have a
distinct impact on constraining $\PM$, which we will discuss in \sref{sec:discussion}.

In comparison, the $\FP$ MF never exceeds CDM and drops significantly for
$\Mvir \lesssim 3\times10^{10} \Msun$. Moreover, unlike $\EA$, the ratio
between the $\FP$ and CDM MFs is found to be almost redshift-independent,
in agreement with the previous study \citep{Schive2016}.

Particle simulations with an initial power spectrum cut-off
are known to suffer from the formation of low-mass spurious haloes,
mostly confined along cosmic filaments and resulting in an unphysical
upturn at the low-mass end of MF \citep[e.g.,][]{Wang2007}.
However, we do not detect either of these artificial features,
suggesting that the contamination from spurious haloes is minimal. It is
likely because the spurious haloes are more prominent for
$\Mvir \lesssim 10^9 \Msun$ with $\PM \sim 1$ \citep{Schive2016}, which is
beyond the minimum halo mass adopted here.

\subsection{Halo Assembly History}
\label{subsec:halo_assembly}

Figs \ref{fig:proj_dens} -- \ref{fig:mass_func} suggest very different
assembly histories between massive $\psiDM$ and CDM haloes, which we detail below.
We select all haloes more massive than $10^{12} \Msun$ at $z=4$, leading to
$\sim 170$ candidates in each model, and trace their progenitors to $z=11$. We define
major mergers as those with progenitor mass ratio above $1:3$.

\begin{figure}
\centering
\includegraphics[width=\columnwidth]{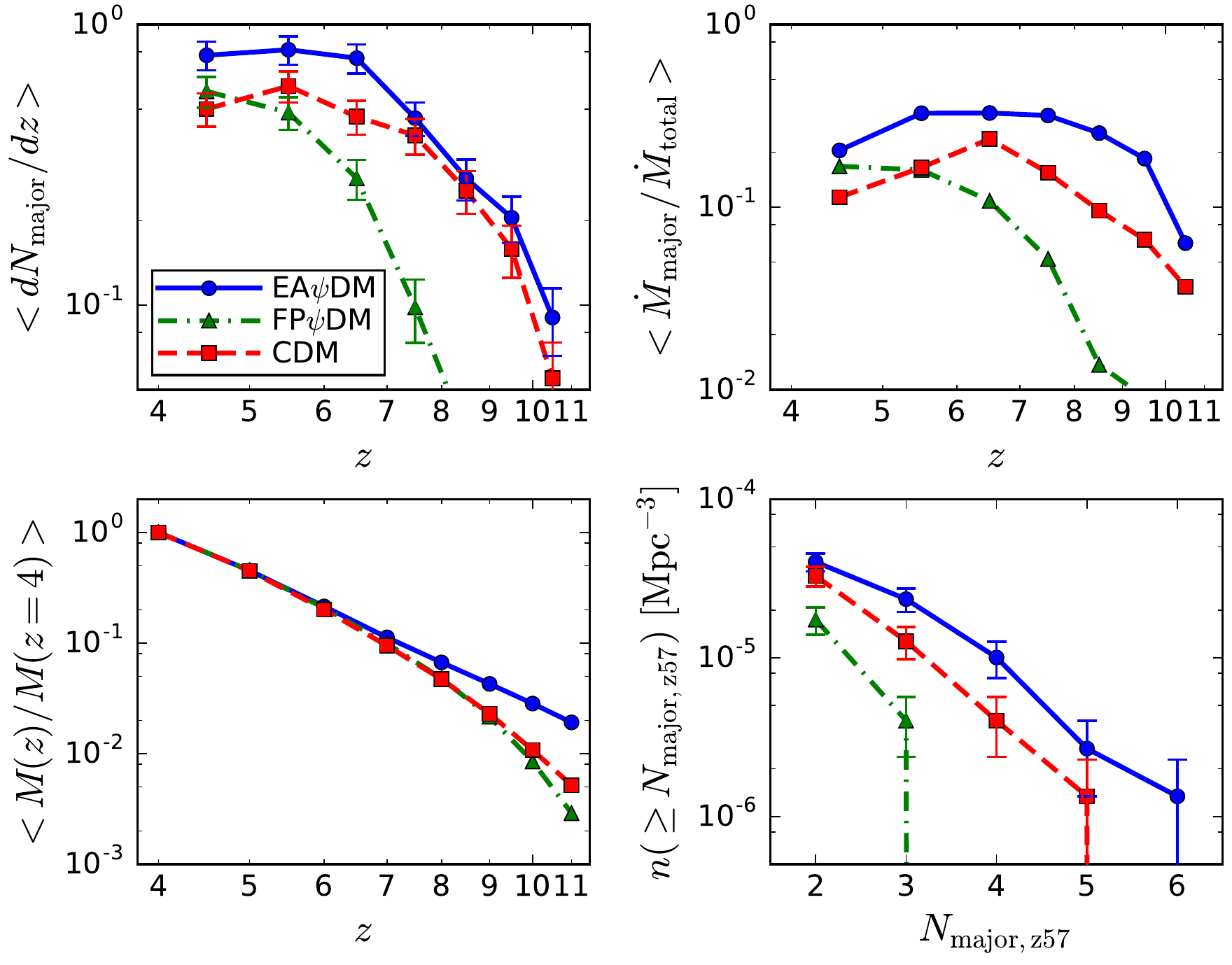}
\caption{
Halo assembly history at $4 \le z \le 11$:
the mean major merger rate per halo per unit $z$ (upper left),
the fraction of mass accretion rate via major merger (upper right),
the mean mass assembly history (lower left),
and the cumulative halo number density with at least $\Nmajor$
major mergers at $5 \le z < 7$ and at most one major merger at $4 \le z < 5$
(lower right). Error bars are Poisson counting uncertainties.
See text for details.
}
\label{fig:mass_assembly}
\end{figure}

\fref{fig:mass_assembly} shows various aspects of the halo assembly history.
The upper left panel shows the mean major merger rate per halo per unit $z$.
The most striking feature in $\EA$ is the apparently higher major merger rate
at $4 \lesssim z \lesssim 7$, exceeding CDM by $\sim 50\%$, followed
by a sharp transition at $z \sim 7$. This feature is verified to be insensitive
to the adopted progenitor mass ratio. In comparison, $\FP$ exhibits a much lower
merger rate at higher redshifts due to the strong suppression of
low-mass haloes. The upper right panel shows the fraction of mass accretion rate
via major mergers. This ratio is found to be less than $\sim 30\%$ in all cases,
suggesting that major mergers do not dominate mass accretion even for $\EA$.

The lower left panel shows the mean mass assembly
history, defined as the halo mass ratio between $z>4$ and $z=4$. Importantly,
the $\EA$ curve is found to be significantly higher
at $z \gtrsim 7$, reaching a factor of $3\text{--}4$ higher than CDM at
$z=10\text{--}11$, while all curves converge at $z \lesssim 7$.
Since all models have similar halo MFs for $\Mvir \gtrsim 10^{12} \Msun$ at
$z=4$ (see \fref{fig:mass_func}), the mean mass assembly fraction shown here
can be regarded as approximately proportional to the average halo mass.
Therefore, it shows that compared to the CDM counterparts, massive $\EA$ haloes
(i) form at higher redshifts, (ii) are a factor or $3\text{--}4$ more massive
at $z=10\text{--}11$, and (iii) experience a slower mass accretion at
$7 \lesssim z \lesssim 11$.

In $\EA$, the findings of a smooth mass accretion at $7 \lesssim z \lesssim 11$ followed
by a steeply rising major merger rate at $z \sim 7$ indicate that a substantial
halo population may experience a rapidly increasing star formation at $z \lesssim 7$.
Interestingly, recent observations show that $\sim 35\%$ of massive
galaxies at $z \sim 4$ are quiescent, with a stellar masses of
$\sim 10^{11} \Msun$, a number density of
$(1.8\pm0.7) \times10^{-5} \Mpc^{-3}$, and an exceedingly efficient star
formation at $5 \lesssim z \lesssim 7$ \citep{Straatman2014, Glazebrook2017},
although still under debate \citep{Simpson2017}.
These features cannot be easily explained by current CDM simulations
\citep[e.g.,][]{Rodriguez2016}.
To investigate this issue in the context of $\EA$, we show in the lower right
panel of \fref{fig:mass_assembly} the cumulative halo number density
with \emph{at least} $\Nmajor$ major mergers at $5 \le z < 7$ and \emph{at most}
one major merger at $4 \le z < 5$.
Intriguingly, we find that $\EA$ haloes have a noticeably higher number
density of these extreme events, exceeding CDM by factors of 2.5 and 2
for $\Nmajor \ge 4$ and $5$, respectively, and
reaching $(1.0\pm0.3) \times10^{-5} \Mpc^{-3}$ for
$\Nmajor \ge 4$. It suggests that compared to CDM, massive $\EA$ haloes
may exhibit more prominent starbursts at $5 \lesssim z \lesssim 7$.

Note, however, that the CDM simulations of \citet{Rodriguez2016} have a
box size of as large as $\sim 100^3 \Mpc^3$, implying that
$\EA$ may not completely solve the puzzle since it only increases the number of
candidates by a factor of $2 \text{--} 3$. Moreover, $\sim 50\%$ of $\EA$
haloes with $\Nmajor \ge 4$ shown in \fref{fig:mass_assembly}
still undergo one major merger at $4 \le z < 5$ and thus may not be fully quiescent.
Larger hydrodynamical simulations,
ideally coupled with dynamical quantum effect, are necessary for addressing
this subject in more detail.

\section{Discussion and Summary}
\label{sec:discussion}

In this paper, we have conducted cosmological simulations to compare the
halo mass functions (MF) and assembly histories between three different
dark matter models, namely, the extreme-axion wave dark matter
\citep[$\EA$,][]{ZC2017b}, the free-particle wave dark matter
\citep[$\FP$,][]{Hu2000,Schive2014a}, and the cold dark matter (CDM).
Both $\psiDM$ models feature a strong suppression of low-mass haloes,
the scale of which is mainly determined by the dark matter particle
mass ($\PM$). However, the $\EA$ model introduces a second free parameter,
the initial field angle $\initangle$, which, for the same $\PM$, can result
in a dark matter transfer function with a larger cut-off wavenumber and a broad
spectral bump before the cut-off (see \fref{fig:power_spec}). Both features
are expected to produce more abundant low-mass haloes and significantly alter
the halo formation history. The main motivation of this work is to quantify
these effects.

Our major results can be summarized as follows.
\begin{itemize}
\item{
$\EA$ MF outnumbers CDM in a wide mass range peaking at
$\Mvir \sim 3\times10^{10} \Msun$. The MF excess is more prominent at higher
redshifts, reaching a factor of $2 \text{--} 3$ at $z \sim 4$ and exceeding
an order of magnitude at $z \sim 10$ (see \fref{fig:mass_func}).
}
\item{
$\EA$ MF at $\Mvir \lesssim 3\times10^{10}$ is in excess of $\FP$ by one and
two orders of magnitude at $z \sim 4$ and $z \sim 10$, respectively
(see \fref{fig:mass_func})}.
\item{
Compared to the CDM counterparts, massive $\EA$ haloes are on average
$3\text{--}4$ times more massive at $z=10\text{--}11$ due to their earlier
formation, and then undergo a slower mass accretion at
$z \gtrsim 7$. Afterward, their mean major merger rate rises sharply and
exceeds CDM by $\sim 50\%$ at $4 \lesssim z \lesssim 7$
(see the upper and lower left panels of \fref{fig:mass_assembly}),
suggesting more prominent starbursts in $\EA$ haloes at this epoch.
}
\end{itemize}

The finding of substantially more low-mass haloes in $\EA$ compared to $\FP$
with the same $\PM$ may have a distinct impact on constraining $\PM$.
If we naively estimate an effective particle mass $\PMeff$ of $\EA$
by equating the cumulative halo MFs,
$n_{\EA}({\ge}M_{\rm min},\PM) = n_{\FP}({\ge}M_{\rm min},\PMeff)$
with $\PM \sim 1$ and an arbitrarily small $M_{\rm min}$ (e.g., $\sim10^8 \Msun$),
where $n_{\FP}$ is estimated using Eq. (7) of \citet{Schive2016},
we obtain $\PMeff \sim 5 \PM$ at $z=4$. Since the core radii of $\psiDM$ haloes
are determined by $\PM$ instead of $\PMeff$, it thus suggests that $\EA$
could reduce the tension in $\PM$ required by the Ly$\alpha$ forest
\citep[$\PM \gtrsim 10$,][]{Irsic2017, Armengaud2017},
the high-$z$ luminosity functions and reionization
\citep[$\PM \gtrsim 1$, e.g.,][]{Schive2016, Corasaniti2017},
and the kpc-scale cores of dwarf spheroidal galaxies
\citep[$\PM \sim 1$, e.g., ][]{Schive2014a, Chen2017}.
Moreover, note that this definition of $\PMeff$ only considers the halo
number density and disregards the fact that $\EA$ haloes are on average
more massive, and thus we should regard it as a conservative lower limit
in this sense.
Also, the choice of $\initangle=0.2^{\circ}$ in this work is to some degree arbitrary
to demonstrate the effect of the $\EA$ model, and a somewhat smaller $\initangle$
may alleviate this tension further. Regardless of this possibility,
$\psiDM$ simulations coupled with both dynamical quantum effect and baryons
are essential for a more quantitative study on this subject.

\section{Acknowledgement}
\label{sec:acknowledgement}

We thank Ui-Han Zhang for providing the initial $\EA$ power spectrum.
The simulations were conducted on the Campus Cluster at the University
of Illinois at Urbana-Champaign.
This work is supported in part by MOST of Taiwan under the grant MOST 103-2112-M-002-020-MY3.

\bibliographystyle{mnras}
\bibliography{ref}

\bsp
\label{lastpage}
\end{document}